\documentclass[runningheads]{svmult}
\usepackage{makeidx}   % allows index generation
\usepackage{graphicx}  % standard LaTeX graphics tool
\usepackage{subeqnar}  % subnumbers individual equations
\usepackage{multicol}  % used for the two-column index
\usepackage{cropmark} % cropmarks for pages without
\usepackage{physprbb}  % centered layout of diverse elements, etc.
\def\etal{{\it et al}}
\def\AEF{A.E. Faraggi}
\def\NPB#1#2#3{Nucl. Phys. B \textbf{#1},  #3 (#2)}
\def\PLB#1#2#3{Phys. Lett. B \textbf{#1},  #3 (#2)}

\def\PRD#1#2#3{Phys. Rev. D  \textbf{#1},  #3 (#2)}

\def\PRL#1#2#3{Phys. Rev. Lett. \textbf{#1},  #3 (#2)}
\def\PRT#1#2#3{Phys. Rep. {\textbf#1},  #3 (#2)}
\def\MODA#1#2#3{Mod. Phys. Lett. A {\textbf #1},  #3 (#2)}
\def\IJMP#1#2#3{Int. J. Mod. Phys. A {\textbf #1}, #3 (#2)}

\def\APJ#1#2#3{Astrophys. J. {\textbf #1}, #3 (#2)}

\begin{document}
\title*{Superstring Phenomenology
\newline  in Light of LEP, KamLAND and WMAP}
\toctitle{Superstring Phenomenology
\protect\newline in Light of LEP, KamLAND and WMAP}
% allows explicit linebreak for the table of content
%
%
\titlerunning{Superstring Phenomenology}
% allows abbreviation of title, if the full title is too long
% to fit in the running head
%
\author{Alon~E.~Faraggi\inst{1}}
\authorrunning{Alon E. Faraggi}
% if there are more than two authors,
% please abbreviate author list for running head
%
%
\institute{Theoretical Physics Department, University of Oxford, 
Oxford OX1 3NP, UK}

\maketitle              % typesets the title of the contribution

\begin{abstract}
The experimental data of the past decade suggests that the quantum
gravity vacuum should possess two key ingredients. The existence of
three generations and their embedding in $SO(10)$ representations. 
The $Z_2\times Z_2$ orbifold of the heterotic string
provides examples of vacua that accommodate these properties.
The utilization of string dualities to explore these models
is discussed. Classification of the $Z_2\times Z_2$ orbifold
with geometric shifts on complex manifolds demonstrates that three
generations are not obtained solely with symmetric shifts on complex
tori, but necessarily utilize an asymmetric shift or a nonperturbative
breaking of the GUT gauge group. The issue of mass and mixing spectrum
in the neutrino versus the quark sector is examined. 
\end{abstract}

\section{Introduction}
\subsection{Experimental guidelines}
The passing decade provided remarkable amount of high quality
experimental data, that deeply affects our perception of
physical reality. The collider experiments, in particular 
the LEP collaboration at CERN \cite{lep}, confirmed the validity of
the Standard Model to a degree precedented only by quantum
electrodynamics. The precision measurements at LEP also
yielded one of the vital experimental clues for physics 
beyond the Standard Model. By restricting the number 
of light left--handed neutrinos below the $Z$--threshold
to $N_\nu\sim3$, LEP has constrained the number
of light chiral generations to be three and only three!
This observation is one of the vital clues in the 
quest for the fundamental unification of the matter and
interactions. In the course of the previous decade the 
impressive output of the LEP collaboration continued
relentlessly, providing for the first time solid 
confirmation of the three and four gauge boson vertices,
hence confirming the non--Abelian character of the
electroweak interactions. The wealth of experimental
data from LEP is a triumph of scientific indulgence. 

In the passing five years the neutrino observatories produced revealing new
data on the neutrino sector of the Standard Model \cite{superk,sno,kamland},
with profound implications on its fundamental origins.
This new data represents
the pinnacle of accumulative and steady progress
in the experimental detection of neutrinos
over the past fifty years. 
For the first time it provides conclusive 
evidence that the neutrinos are massive and that the 
Standard Model spectrum is augmented by the right--handed
neutrinos. While this eventuality has been anticipated since 
the early seventies \cite{patisalam}, one should note the profound
difference between naive theoretical expectations versus
the experimentally confirmed observations. 

In the past year the WMAP collaboration \cite{wmap} provided
unprecedented accurate measurements of the cosmic microwave
background radiation. The new data impose further constraints
on the range of allowed neutrino masses. But, perhaps more importantly,
it brings cosmological observations closer to providing
valuable quantitative data that may prove to be revealing
on the physical properties of the universe. In the
least it provides further accurate data on the energy
composition of the universe. Perhaps more
intriguing is the, yet statistically insignificant,
indications of correlations at large angles. If proved correct
these will have unprecedented implications on the structure of
the universe. In particular on its potential realization
in the framework of string theory.

Lastly, the forthcoming decade promises to be even grander,
with AUGER, PLANCK, LHC, The Linear Collider (TLC), and other experiments.

\subsection{Theoretical considerations}
The augmentation of the Standard Model by the right--handed neutrinos
has profound implications on its fundamental origin.
It provides strong support for the embedding of the
Standard Model spectrum, generation by generation,
in spinorial 16 representation of $SO(10)$.
Furthermore, the neutrino data from cosmic and
reactor experiments indicate surprising new structures
that go beyond the conventional GUT expectations. 
The observational data from collider facilities and cosmic observatories
indicates two pivotal ingredients in the quest to understand the fundamental
origins of the Standard Model. The existence of three generations
and their embedding, generation by generation, in spinorial 16
representations of $SO(10)$. In addition to the gauge sector
the Standard Model data consists of the flavor sector. While the 
gauge observables find their natural origins in Grand Unified Theories,
the flavor observables do not. This implies that fundamental 
understanding of the flavor parameters must arise from a theory
that unifies the gauge and gravitational interactions. 
String theory provides a consistent contemporary framework
for perturbative quantum gravity. 
As such
string theory produces viable machinery to study how 
a fundamental theory of quantum gravity may determine the 
Standard Model parameters. 

Additionally the consistency of string theory requires some
additional structures beyond the Standard Model.
These arise in the form of extra dimensions, supersymmetry
and additional matter and gauge states beyond the SM spectrum.
The question, however, is whether these additional
structures yield any experimental imprints that
may be observed.
Since the main collider experiments in the forthcoming decades
will target the physics at the electroweak scale, and the 
associated Higgs mass generation mechanism, it is hoped
that at least some of these structures will be accessible in
such experiments. These may come in the form of low scale 
supersymmetry, or large extra dimensions that arise in some
limits of the underlying string theory. Additionally,
proton decay experiments and cosmic observatories may
provide experimental probes to the physics at larger scales. 

The utility of string theory in my view is to study
how the parameters of the Standard Model may arise
from a theory of quantum gravity. In the past few years
a new view of string theories has been developed \cite{Mtheoryreviews}.
In the framework of M--theory the five different string theories
in ten dimensions, together with eleven dimensional supergravity
are seen to be limits of a more fundamental theory, traditionally
dubbed M--theory. We should therefore question how this new
picture affects the utilization of string theory toward
phenomenological studies. In this respect we should regard 
the perturbative string limits as effective limits, non
of which can fully characterize the true vacuum, which
should possess some non--perturbative realization. 
The new picture of M--theory suggests that the effective
limits may at best probe some features of the non--perturbative
vacuum, and that different limits may be instrumental to 
extract different properties. For example, 
the $SO(10)$ embedding of the Standard Model spectrum can only be seen
in the heterotic limit because this is the only limit that gives
rise to the chiral 16 of $SO(10)$ in the effective low energy field
theory. On the other hand, in the perturbative heterotic--string limit
the dilaton, whose VEV governs the string gauge and gravitational
couplings, has a run--away potential and cannot
be stabilized at a finite value. However, we should regard 
the heterotic limit as the zero coupling expansion of the more
basic theory. With our 
present understanding of string theories in the context
of their M--theory embedding it is clear that we should not
in fact expect the dilaton to be stabilized in the heterotic limit. 
In order to stabilize the dilaton we have to move away from
the zero coupling expansion, or to move away from the 
perturbative heterotic--string limit. The existence
of the classical eleven dimensional limit in which the dilaton
is interpreted as the moduli of the eleventh dimension
lends credence to this general expectation. Thus, the issue
of dilaton stabilization may be more accessible, even if not yet
fully resolved, in other limits of the underlying theory,
rather than in the perturbative heterotic string limit.

The new M--theory picture therefore suggests the following approach
toward utilization of string theory for phenomenological studies.
Suppose that in some effective limit we are able to identify
a string vacuum that exhibits viable phenomenological characteristics.
This would entail identifying a particular class of string
compactifications on which the effective string theory is compactified.
The new picture of M--theory then suggests that additional information
on the properties of the non--perturbative vacuum may be gleaned
by compactifying other string limits on the same class of manifolds.

In this talk I will discuss the efforts to develop such an approach
undertaken by the string phenomenology group at Oxford. Additionally
I will discuss recent work the geometrical correspondence
of the three generation free fermionic models. These models explain
the origin of the three generations as arising from the three twisted
sectors of the $Z_2\times Z_2$ orbifold compactification. 
Hence the improved understanding of the geometrical correspondence
of the free fermionic models aims to develop a geometrical explanation
for the origin of the three generations. 
I discuss the potential implications
of the large neutrino mixing angles for string and beyond the
Standard Model phenomenology, and recent efforts aimed at developing the
tools to utilize the forthcoming Ultra High Energy Cosmic Ray (UHECR)
experiments to probe physics beyond the Standard Model and string theory.

\section{High versus low}\label{membedding}

To illustrate the approach toward string phenomenology advocated
in this paper we have to make some judicial assumptions in regard to
the physics beyond the Standard Model. There are basically two
orthogonal directions that may be pursued. The first assumes
that the Standard Model remains perturbative up to a very
high scale, and that the ultra--violet cut--off, set by quantum
gravity, is at the Planck scale. The second essentially assume that 
the Standard Model breaks down already not far beyond the electroweak
scale and that the ultra--violet non--perturbative cut--off
is at the TeV, or multi--TeV scale. In this paper the first
possibility is pursued. Aside from the elegant $SO(10)$ embedding
of the Standard Model spectrum, this case is also supported by the
viability of logarithmic running of the Standard Model gauge and
matter sectors parameters, whereas the scalar sector requires
the introduction of supersymmetry. Proton longevity and suppression
of left--handed neutrino masses lends additional support to this
picture. 

Thus, the basic properties that we would like our string vacuum to possess
are the existence of three chiral generations and their embedding in 
$SO(10)$ representations. 
A class of string models that yield these properties are the
three generation free fermionic models. The structure of these
models and related phenomenological studies have been amply discussed
in the past and therefore in this paper only a brief discussion will
be given.

\section{Dualities}
String theory exhibits various forms of dualities, {\it i.e.} 
relation between different theories at large and small radii
of the compactified manifold and at strong and weak coupling. 
The first type is the T--duality \cite{tduality}. Consider a point particle 
moving on a compactified dimension $X$, which obeys the condition
$X\sim X +2\pi R m$. Single valuedness of the wave function of the 
point particle $\Psi\sim {\rm Exp}(iP X )$ implies that the momenta
in the compact direction is quantized $P={m\over R}$ with $m\in Z$. 
Now consider a string moving in the compactified direction.
In this case the string can wrap around the compactified dimension
and produce stable winding modes. Hence the left and right--moving
momenta in the case of the closed string are given by
$$P_{L,R}={m\over R}\pm {{nR}\over \alpha^\prime}$$
and the mass of the string states is given by
$${\rm mass}^2=\left({n\over R}\right)^2+\left({{m R}\over
\alpha^\prime}\right)^2$$
this is invariant under exchange of large and small radius together 
with the exchange of winding and momentum modes, {\it i.e.}
$${1\over R}\leftrightarrow {R\over\alpha^\prime}~~~{\rm with}~~~
m\leftrightarrow n$$
and is an exact symmetry in string perturbation theory. Furthermore, 
there exist the self--dual point, 
$$ R={\alpha^\prime\over R},$$
which is the symmetry point 
under T--duality. In the case of the supersymmetric string on a 
compactified coordinate the T--duality operation interchanges 
\begin{eqnarray}
{\rm type ~IIA} & ~\Leftrightarrow~ & {\rm type~ IIB}\\
{\rm Heterotic~} SO(32) & ~\Leftrightarrow~ & {\rm Heterotic}~ E_8\times E_8
\end{eqnarray}
Now, all this
is of course well known since the late 80's. However, the following point
is not well appreciated. It is also well known that for specific
values of its  radius, the compactified coordinate can be realized 
as specific rational conformal field theories propagating on
the string world--sheet. In particular, there exist such a value for
which a compactified coordinate can be represented in terms
of two free Majorana--Weyl fermions. It so happens that, in some
normalization, the self--dual point is at $R=1/\sqrt2$ whereas the
free fermionic point is at $R=1$. Hence, the two points do not overlap
and the free fermionic point does not coincide with the self--dual point \cite{ginsparg}. 
However, this is merely an artifact of the fact that we have been talking
here about bosonic string. In the case of the supersymmetric string the 
two points, in fact, do coincide. This is a remarkable observation for
the following reason. While we do not yet know at what value the compactified
coordinate are fixed, naively we would expect that they are stabilized
around a symmetry point or at infinity. The self--dual point under T--duality
is precisely such a symmetry point. Hence, near the self--dual point,
which is the symmetry point under T--duality and around which
we may expect that the compactified dimensions stabilize, we can represent
the compact dimension as a pair of free Majorana--Weyl fermions
propagating on the string world--sheet. Of course, the real picture 
may be much more complicated. But a a first approximation this is the 
naive expectation, based on the symmetry properties of string theory.

T--duality is perturbative and exhibit itself in the exchange of
the spectrum and the superpotential. Thus, it can be checked order by order in
perturbation theory. In the past decade significant progress
in understanding duality symmetries which are nonperturbative,
{\it i.e.}
that exchange weak with the strong coupling, has been achieved.
The starting point in this program
was the Seiberg-Witten solution of $N=2$
supersymmetric pure $SU(2)$ gauge theory \cite{sw}. 
In the supersymmetric theory the gauge coupling is extended to a complex 
parameter $\tau=\theta/2\pi+i4\pi/g^2$ where $\theta$
is the axial coupling and 
$g$ is the field strength coupling. The strong-weak
duality extends to a $SL(2,Z)$ transformations of the parameter $\tau$. 
In the Seiberg-Witten solution the exact vacuum structure of the theory
is parameterized in terms of a genus one Riemann surface. 

In string theory we have naively a similar situation. The 
gauge coupling is fixed by the VEV of the dilaton field. 
The dilaton field, combined with the space--time components of
the antisymmetric tensor field forms
a modular parameter. In M--theory \cite{Mtheoryreviews}
this complex field is identified with the moduli field of a new dimension 
and hence the $SL(2,Z)$ symmetry of this moduli field translates into
a duality which exchanges strong and weak coupling \cite{sswd,wmtheory}.
The different string limits are related under the
strong--weak coupling exchange, and by T--duality after compactification
to a lower dimension. The question that we examine in the work reported 
here is how to utilize this novel understanding of string theory 
for phenomenological studies.  
\section{Realistic string models}\label{rsm}
As discussed above the key properties that we would like our string
vacuum to possess are the existence of three generations and their 
embedding in $SO(10)$ representations. The only effective string limit
that preserves the $SO(10)$ embedding is the heterotic limit, because
this is the only limit that produces the chiral 16 representation
of $SO(10)$ in the perturbative spectrum. To build realistic models
we compactify the ten dimensional heterotic string to the heterotic string
in four dimensions. This is achieved by choosing a six
dimensional manifold on which the string theory is compactified.
Typically, the six dimensional internal manifolds can be represented 
as toroidal orbifolds. A $D$--dimensional tori is represented in terms
of a $D$--dimensional Euclidean space, modded by a lattice translation. 
The orbifold is in turn obtained by moding the $D$--dimensional
torus by an internal symmetry. Thus, for example, the circle
is one dimensional Euclidean space modded by $2\pi R$ identification
under translation, and its $Z_2$ orbifold, which is a line segment with
two fixed points, is obtained by identifying points by reflection
across the real axis. 

The $Z_2$ orbifold plays a crucial role in the discussion to follow. 
The three generation free fermionic models correspond to
$Z_2\times Z_2$ orbifold of a six dimensional torus. The first 
$Z_2$ acts on the first four coordinates, whereas the second 
acts on the last four. The $Z_2\times Z_2$ orbifold models, 
through their realization in the free fermionic formulation, 
produce three generation models with $SO(10)$ embedding.
Furthermore, the structure of the $Z_2\times Z_2$ orbifold naturally
correlates the existence of three generation in nature
with the underlying geometry. This arises due to the fact that 
each the $Z_2\times Z_2$ orbifold has exactly three twisted sectors. 
Each of the light chiral generations then arises from a distinct 
twisted sector. Hence, in these models the existence of three generations
in nature is seen to arise due to the fact that we are dividing
a six dimensional compactified manifold into factors of 2. In simplified
terms, three generations is an artifact of 
$${6\over 2}~~=~~1~+~1~+~1$$
One may further ask whether there is a reason that the $Z_2$ 
orbifold would be preferred versus higher orbifolds.
Previously we argued that the free fermionic point 
coincides with the self--dual point under $T$--duality,
which is where we would naively expect the compactified 
dimensions to stabilize. The special property of the 
$Z_2$ orbifold that sets it apart from higher orbifolds,
is the fact that the $Z_2$ orbifold is the only one that acts
on the coordinates as real coordinates, rather than complex 
coordinates. Whether this property plays a role in the string 
vacuum selection is yet to be understood.

\section{Free fermionic model building}\label{ffmb}
The three generation $Z_2\times Z_2$ orbifold models were studied
in the free fermionic formulation \cite{fff}. These models were reviewed
in the past in these conference series \cite{btd979902}, and
I therefore give here only a
brief summary. The models are constructed in terms of a set of boundary
condition basis vectors that define the transformation properties 
of the 20 left--moving and 44 right--moving real fermions
around the noncontractible loops of the one--loop vacuum to vacuum 
amplitude. 

The first five basis vectors of the realistic free fermionic
models consist of the NAHE set \cite{nahe}.
The gauge group after the NAHE set is
$SO(10)\times E_8\times SO(6)^3$ with $N=1$ space--time supersymmetry,
and 48 spinorial $16$ of $SO(10)$, sixteen from each sector $b_1$,
$b_2$ and $b_3$. The three sectors $b_1$, $b_2$ and $b_3$ are
the three twisted sectors of the corresponding $Z_2\times Z_2$  
orbifold compactification. The $Z_2\times Z_2$ orbifold is special
precisely because of the existence of three twisted sectors,
with a permutation symmetry with respect to the horizontal $SO(6)^3$
charges.

The NAHE set is common to a large class of three generation
free fermionic models. The construction proceeds by adding to the
NAHE set three additional boundary condition basis vectors
which break $SO(10)$ to one of its subgroups: $SU(5)\times U(1)$
\cite{revamp}, $SO(6)\times SO(4)$ \cite{patisalamstrings},
$SU(3)\times SU(2)\times U(1)^2$ \cite{fny,eu,top,cfn},
or $SU(3)\times U(1)\times SO(4)$ \cite{lrsstringmodels}.   
At the same time the number of generations is reduced to
three, one from each of the sectors $b_1$, $b_2$ and $b_3$.
The various three generation models differ in their
detailed phenomenological properties. However, many of  
their characteristics can be traced back to the underlying
NAHE set structure. One such important property to note
is the fact that as the generations are obtained  
from the three twisted sectors $b_1$, $b_2$ and $b_3$,
they automatically possess the Standard $SO(10)$ embedding.
Consequently the weak hypercharge, which arises as
the usual combination $U(1)_Y=1/2 U(1)_{B-L}+ U(1)_{T_{3_R}}$,
has the standard $SO(10)$ embedding.

The massless spectrum of the realistic free fermionic models
then generically contains three generations from the
three twisted sectors $b_1$, $b_2$ and $b_3$, which are
charged under the horizontal symmetries. The Higgs spectrum
consists of three pairs of electroweak doublets from the
Neveu--Schwarz sector plus possibly additional one or
two pairs from a combination of the two basis vectors  
which extend the NAHE set. Additionally the models
contain a number of $SO(10)$ singlets which are
charged under the horizontal symmetries and 
a number of exotic states.

Exotic states
arise from the basis vectors which extend the NAHE
set and break the $SO(10)$ symmetry \cite{ccf}. Consequently, they
carry either fractional $U(1)_Y$ or $U(1)_{Z^\prime}$ charge.
Such states are generic in superstring models
and impose severe constraints on their validity.
In some cases the exotic fractionally charged
states cannot decouple from the massless
spectrum, and their presence invalidates otherwise
viable models \cite{otherrsm,penn}.
In the NAHE based models the fractionally
charged states always appear in vector--like
representations. Therefore, in general mass
terms are generated from renormalizable or nonrenormalizable
terms in the superpotential. However, the mass terms which arise
from non--renormalizable terms will in general be suppressed,
in which case the fractionally charged states may have
intermediate scale masses.
The analysis of ref. \cite{cfn} demonstrated the  
existence of free fermionic models with solely the 
MSSM spectrum in the low energy effective field theory of the
Standard Model charged matter.
In general, unlike the ``standard'' spectrum, the ``exotic'' spectrum is
highly model dependent.

\section{Phenomenological studies of free fermionic models}

I summarize here some of the highlights of the phenomenological studies
of the free fermionic models. This
demonstrates that the free fermionic string models indeed
provide the arena for exploring many the questions relevant 
for the phenomenology of the Standard Model and Unification.
The lesson that should be extracted is that
the underlying structure of these models, generated   
by the NAHE set, produces the right features for
obtaining realistic phenomenology. It provides further
evidence for the assertion that the true string   
vacuum is connected to the $Z_2\times Z_2$ orbifold
in the vicinity of the free fermionic point in the
Narain moduli space. Many
of the important issues relating to the phenomenology of
the Standard Model and supersymmetric unification have been
discussed in the past in several prototype free fermionic
heterotic string models. These studies have been reviewed in
the past and I refer to the original literature and additional
review references \cite{review,btd979902}.
These include among others: top quark mass prediction \cite{top}, several
years prior to the actual observation by the CDF/D0 collaborations
\cite{cdfd0};
generations mass hierarchy \cite{NRT}; CKM mixing \cite{CKM};
superstring see--saw mechanism \cite{seesaw}; Gauge coupling
unification \cite{gcu}; Proton stability \cite{ps};
supersymmetry breaking and squark degeneracy \cite{fp2,dedes}.
Additionally,
it was demonstrated in ref. \cite{cfn} that at low energies the model
of ref. \cite{fny}, which may be viewed as a
prototype example of a realistic free fermionic model,
produces in the observable sector solely the MSSM charged spectrum.
Therefore, the model of ref. \cite{fny}, supplemented
with the flat F and D solutions of ref. \cite{cfn}, provides
the first examples in the literature of a string model
with solely the MSSM charged spectrum
below the string scale. Thus, for the first time it provides
an example of a long--sought Minimal Superstring Standard Model!
We have therefore identified
a neighborhood in string moduli space which is potentially
relevant for low energy phenomenology. While we can suggest  
arguments, based on target--space duality considerations why this
neighborhood may be selected, we cannot credibly argue that
similar results cannot be obtained in other regions
of the string moduli space. Nevertheless, the results summarized
here provide the justification for further
explorations of the free fermionic models. Furthermore,
they provide motivation to study these models
in the nonperturbative context of M--theory.
In this context the basis for our studies is the
connection of the free fermionic models with the
$Z_2\times Z_2$ orbifold, to which I turn in section \ref{z2z2orbifold}.

I would like to emphasize that it is not suggested that any of the
realistic free fermionic models is the true vacuum of our world.
Indeed such a claim would be folly. Each of the phenomenological
free fermionic models has its shortcomings, that if time and space
would have allowed could have been detailed. While in principle
the phenomenology of each of these models may be improved
by further detailed analysis of supersymmetric flat directions,
it is not necessarily the most interesting avenue for exploration.
The aim of the studies outlined above is to demonstrate that
all of the major issues, pertaining to the phenomenology of the
Standard Model and unification, can in principle be
addressed in the framework of the free fermionic models,
rather than to find the explicit solution that accommodates
all of these requirements simultaneously. The reason being
that even within this space of solutions there is till a vast
number of possibilities, and we lack the guide to select the
most promising one. What is being proposed is that these
phenomenological studies suggest that the true string vacuum
may share some of the gross structure of the free fermionic models.
Namely, it will possess the structure of the $Z_2\times Z_2$
orbifold in the vicinity of the free fermionic point in the
Narain moduli space. This perspective provides the motivation  
for the continued interest in the detailed study of this gross structure,
and specifically in the framework of M--theory, as discussed below.

The free fermionic models also serve as a laboratory to study 
possible signatures beyond the Standard Model. Perhaps most fascinating among
those is the existence of exotic matter states \cite{fccp,ccf}
that can lead to to experimental signatures in the form of
energetic neutrinos from the sun \cite{fop}, or in the form
of candidates for dark matter and top--down UHECR scenarios \cite{cfp}.
The later is particularly exciting due to the forthcoming Pierre Auger and
EUSO experiments that will provide more statistics on UHECR. 

\section{Correspondence with $Z_2\times Z_2$ orbifold}\label{z2z2orbifold}
The key property of the free fermionic models that is sought
in the attempts to investigate these models in the nonperturbative 
framework of M--theory is their relation to the $Z_2\times Z_2$ 
orbifold. In the fermionic language the models are defined 
in terms of the boundary condition basis vectors. Extending the 
NAHE set with the additional basis vector $\xi_2$, we can regenerate
the model defined by this set by using instead the set
$\{{\bf1 },S,\xi_1,\xi_2,b_1,b_2\}$. Here the vacuum produced
by the four basis vectors $\{{\bf 1},S,\xi_1,\xi_2\}$
is an $N=4$ supersymmetric toroidal compactification
with $SO(12)\times E_8\times E_8$ right--moving gauge 
group. Adding the two basis vectors $b_1$ and $b_2$ 
correspond to the $Z_2\times Z_2$ action and produces a
vacuum with $SO(4)^3\times E_6\times U(1)^2\times E_8$
and 27 generations plus 3 anti--generations of the 27
representation of $E_6$. The same model is obtained 
in the bosonic language by specifying the VEVs of the background
fields that produce the enhanced $SO(12)$ lattice
and moding by the $Z_2\times Z_2$ orbifold projections \cite{foc}.
The resulting manifold has $(h_{1,1},h_{2,1})=(27,3)$
and is referred to as $X_2$.

The $Z_2\times Z_2$ orbifold at the free fermionic point in the 
Narain moduli space hence produces a vacuum with a net number of 
24 generations. However, the $Z_2\times Z_2$ orbifold at a generic
point in the moduli space produces
a model with $(h_{1,1},h_{2,1})=(51,3)$ or a net number
of 48 generations. I denote this manifold as $X_1$.
Hence, there is a discrepancy by a factor of 2
between the two models. This discrepancy is in fact
crucial both from the point of view of pursuing the M--embedding of
these models as well as trying to understand the origin of the three 
generations. Below we expand on these aspects.

For our purpose here it is important to observe that the
two manifolds, $X_1$ and $X_2$, may be connected
by adding a freely acting twist or shift.
Let us first start with the compactified
$T^1_2\times T^2_2\times T^3_2$ torus parameterized by  
three complex coordinates $z_1$, $z_2$ and $z_3$,
with the identification
\begin{equation}
z_i=z_i + 1\,, \qquad z_i=z_i+\tau_i \,,
\label{t2cube}
\end{equation}
where $\tau$ is the complex parameter of each
$T_2$ torus.
With the $Z_2$ identification $z_i\rightarrow-z_i$, a single torus
has four fixed points at
$%$
z_i=\{0,{\textstyle{1\over 2}},{\textstyle{1\over 2}}\,\tau,
{\textstyle{1\over 2}} (1+\tau) \}.
$%$
With the two ${Z}_2$ twists
\begin{eqnarray}
&& \alpha:(z_1,z_2,z_3)\rightarrow(-z_1,-z_2,~~z_3) \,,
\cr
&&  \beta:(z_1,z_2,z_3)\rightarrow(~~z_1,-z_2,-z_3)\,,
\label{alphabeta}
\end{eqnarray}
there are three twisted sectors in this model, $\alpha$,
$\beta$ and $\alpha\beta=\alpha\cdot\beta$, each producing
16 fixed tori, for a total of 48. Adding
to the model generated by the ${Z}_2\times {Z}_2$
twist in (\ref{alphabeta}), the additional freely acting shift
\begin{equation}
\gamma:(z_1,z_2,z_3)\rightarrow(z_1+{\textstyle{1\over2}},z_2+
{\textstyle{1\over2}},z_3+{\textstyle{1\over2}})
\label{gammashift}
\end{equation}
produces again fixed tori from the three
twisted sectors $\alpha$, $\beta$ and $\alpha\beta$.
Under the action of the $\gamma$-shift,
the fixed tori from each twisted sector are paired.
Therefore, $\gamma$ reduces
the total number of fixed tori from the twisted sectors   
by a factor of ${2}$,
yielding $(h_{11},h_{21})=(27,3)$. This model therefore
reproduces the data of the ${Z}_2\times {Z}_2$ orbifold
at the free-fermion point in the Narain moduli space.
The precise form of the shift that reproduces the
$Z_2\times Z_2$ orbifold at the free fermionic model
is given in \cite{partitions}, and differ slightly from
(\ref{gammashift}). However, these two models are in the
same moduli space, and hence are connected by continuous extrapolations.
Eq. (\ref{gammashift}) suffices for our discussion here. 

Despite its innocuous appearance the connection between $X_1$ and $X_2$
by a freely acting shift has an important consequence.
{}From the Standard Model data we may hypothesize
that any realistic string vacuum should possess
at least two ingredients. First, it should contain
three chiral generations, and second, it should
admit their SO(10) embedding. This SO(10) embedding
is not realized in the low energy effective field theory
limit of the string models, but is broken directly at the 
string level. The main phenomenological implication of this
embedding is that the weak-hypercharge has the canonical
GUT embedding.

It has been argued that the ${Z}_2\times {Z}_2$ orbifold naturally
gives rise to three chiral generations. The reason being that
it divides the six dimensional compactified manifold into
three cyclicly symmetric spaces.
It contains three twisted sectors and each of these sectors
produces one chiral generation. The existence of
exactly three twisted sectors arises, essentially, because
we are modding out a three dimensional complex manifold, or
a six dimensional real manifold, by ${Z}_2$ projections,
which preserve the holomorphic three form. Thus, metaphorically
speaking, the reason being that six divided by two equals three. 

However, this argument holds for any ${Z}_2\times {Z}_2$ orbifold
of a six dimensional compactified space, and in particular it 
holds for the $X_1$ manifold. Therefore, we can envision that
this manifold can produce, in principle, models with SO(10) gauge
symmetry, and three chiral generations from the three
twisted sectors. However, the caveat is that this manifold
is simply connected and hence the SO(10) symmetry
cannot be broken by the Hosotani-Wilson symmetry breaking
mechanism \cite{hosotani}.
The consequence of adding the freely acting shift (\ref{gammashift})
is that the new manifold $X_2$, while still admitting
three twisted sectors is not simply connected and hence
allows the breaking of the SO(10) symmetry to one of
its subgroups.

The freely acting shift has the crucial
function of connecting between the simply connected covering manifold
to the non-simply connected manifold. Precisely such a construction
has been utilized in \cite{donagi} to construct non-perturbative
vacua of heterotic M-theory. In the next section I discuss these
phenomenological aspects of M--theory.

\section{M--embeddings}

The profound new understanding of string theory that   
emerged over the past few years means that we can use
any of the perturbative string limits, as well as eleven
dimensional supergravity to probe the properties of
the fundamental M--theory vacuum.
The pivotal property that this vacuum should preserve
is the $SO(10)$ embedding of the Standard Model spectrum.
This inference follows from the fact that also in the
strong coupling limit heterotic M--theory produces
discrete matter and gauge representations. 
Additionally, the underlying compactification should allow
for the breaking of the $SO(10)$ gauge symmetry.
In string theory the prevalent method to break the
$SO(10)$  gauge group is by utilizing Wilson
line symmetry breaking. Compactification of M--theory 
on manifolds with $SU(5)$ GUT gauge group that can broken
to the Standard Model gauge group were discussed in \cite{donagi}.
In \cite{fgi} the analysis was extended to $SO(10)$ GUT gauge
group that can be broken to $SU(5)\times U(1)$. This
work was reviewed in \cite{btd979902} and here I discuss relevant points
for further explorations of the phenomenological free fermionic models.

The key to the construction of ref. \cite{donagi} is the
utilization of elliptically fibered Calabi--Yau threefolds.
These manifolds are represented as a two dimensional
complex base manifold and a one dimensional complex fiber
with a section. 
On these manifolds the equation for the fiber is given 
in the Weierstrass form
$%$
y^2=x^3+f(z_1,z_2)x+g(z_1,z_2)=(x-e_1)(x-e_2)(x-e_3).
$%$
Here $f$ and $g$ are polynomials of
degrees 8 and 12, respectively and are functions of
the base coordinates; $e_1$, $e_2$ and $e_3$ are the three
roots of the cubic equation. Whenever two of the roots coincide
the fiber degenerates into a sphere. Thus, there is a locus
of singular fibers on the base manifold.
These singularities are resolved by splitting the fiber
into two spherical classes $F$ and $F-N$. One being the
original fiber minus the singular locus, and the second 
being the resolving sphere.

A nonperturbative vacuum state of the heterotic M--GUT--theory 
on the observable sector is specified by a set of 
M--theory 5--branes wrapping a holomorphic 2--cycle on the 3--fold. The 
5--branes are described by a 4--form cohomology class $[W]$
satisfying the anomaly--cancellation condition. 
This class is Poincar\'e--dual 
to an effective cohomology class in $H_2(X, {\bf Z})$ that can be written as
$%$
[W]= c_2(TX)-c_2(V_1)-c_2(V_2)=
\sigma_*(w)+c(F-N)+dN,
$%$
where $c_2(TX)$, $c_2(V_1)$ and $c_2(V_2)$ are the 
second Chern classes of the tangent bundle and the two
gauge bundles on the fixed planes; $c,d$ are positive definite
integers, $\omega$ is a class in $B$, and
$\sigma_*(\omega)$ is its pushforward to $X$ under $\sigma$.

The key to the M--theory embedding of the free fermionic models
is their correspondence with the $Z_2\times Z_2$ orbifold. The starting
point toward this end is the $X_1$ embedding manifold with 
$(h_{11},h_{21})=(51,3)$. The manifold is then rendered non--simply
connected by the freely acting involution and the methodology of
ref. \cite{donagi,fgi} can be adopted to construct viable M--theory
vacua. The difference however is that now the fiber is
more singular than the ones previously considered.
The fiber of $X_1$ in Weierstrass form is given by
$%$
y^2 = x^3 + f_8 (w,\tilde w) x z^4 + g_{12}(w,\tilde w) z^6,
$%$
where 
$%$
f_8=\eta-3h^2,~{\rm and}~g_{12}~=~h(\eta-2 h^2),
$%$
$%$
h = K \prod_{i,j=1}^4(w- w_i)(\tilde w - \tilde w_j)
$%$ 
and
$%$
\eta = C \prod_{i,j=1}^4(w- w_i)^2(\tilde w - \tilde w_j)^2.
$%$
Taking $w\rightarrow w_i$ (or ${\tilde w}\rightarrow
{\tilde w}_i$) we have a $D_4$ singular fiber.
These $D_4$ singularities intersect in 16 points, $(w_i,\tilde w_j),\,
i,j=1,\ldots 4$, in the base.
The resolution of the singular fiber in this case
is more involved than the simpler ones previously considered.
It is expected that the richer structure of fiber
classes will yield a richer class of M--theory vacua
with the possibility of new features appearing.

\begin{figure}[b]
\begin{center}
\includegraphics[width=.3\textwidth]{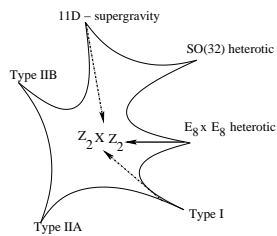}
\end{center}
\caption[]{Phenomenological application of M--theory}
\label{btd02proc1.1}
\end{figure}

Figure (\ref{btd02proc1.1}) illustrates qualitatively the approach
to the phenomenological application of M--theory advocated in this
paper. In this view the different perturbative M--theory
limits are used to probe the properties of a specific
class of compactifications. In this respect one may 
regard the free fermionic models as illustrative examples.
Namely, in the heterotic limit this formulation highlighted
the particular class of models that are connected to the
$Z_2\times Z_2$ orbifold. In order to utilize the M--theory
advances to phenomenological purposes, our task then is
to now explore the compactification of the other
perturbative string limits on the same class of
spaces, with the aim of gaining further insight
into their properties. In this spirit compactifications of 
type I string theory on the $Z_2\times Z_2$ orbifold that are
connected to the free fermionic models have been explored\cite{dave}. 

\section{On the origin of the three generations}

The free fermionic models correspond to $Z_2\times Z_2$
orbifold at an enhanced symmetry point in the Narain moduli space.
As argued above the $Z_2\times Z_2$ orbifold, via its free fermion
realization, naturally produces three generation models arising from the
three twisted sectors. However, the geometrical correspondence
of the free fermionic models is so far understood for the extended
NAHE set models, {\it i.e.} for the case of the $X_2$ manifold 
with 24 generations. Hence, in order to promote the geometrical 
understanding of the origin of the three generations in the free 
fermionic models, it is important to understand the geometrical 
interpretation of the boundary condition basis vectors beyond the 
NAHE set. 

Let us review for this purpose the vacuum structure in the twisted
sectors $b_1$, $b_2$ and $b_3$. In the light--cone gauge
the world--sheet free fermion field
content includes: in the left--moving sector the
two space--time fermions $\psi^\mu_{1,2}$ and the six real triples
$\{\chi_i, y_i,\omega_i\}$ $(i=1,\cdots,6)$; in the right--moving sector
the six real doubles $\{{\bar y}_i,{\bar\omega}_i\}$ $(i=1,\cdots,6)$
and the sixteen complex fermions
$\{{\bar\Psi}^{1,\cdots,5},{\bar\eta}^{1,2,3},{\bar\phi}^{1,\cdots,8}\}$. 
For our purpose the important set is the set of internal real fermions
$\{y,\omega\vert{\bar y},{\bar\omega}\}^{1,\cdots,6}$. We can bosonize
the fermions in this set by defining
$$
{\rm e}^{iX_i}= {1\over\sqrt2}(y_i+i\omega_i)~~~
{\rm e}^{i{\bar X}_i}= {1\over\sqrt2}({\bar y}_i+i{\bar\omega}_i)
$$
We recall that the vacuum of the sectors $b_i$ is made of 12 periodic complex
fermions, $f$, each producing a doubly degenerate vacua
${\vert +\rangle},{\vert -\rangle}$ ,
annihilated by the zero modes $f_0$ and
${{f_0}^*}$ and with fermion numbers  $F(f)=0,-1$, respectively.
The total number of states in each of these sector is therefore
$$2^{12}=\sum_{n=0}^{12}\left(\matrix{12 \cr n\cr}\right).$$
After applying the GSO projections the degeneracy at the level of the 
extended NAHE model distributes as follows:

\begin{eqnarray}
~~~~~~~~~\left[~~y,\omega\vert{\bar y}, {\bar\omega}~~\right]_{b_j}
& \left( \psi^\mu\ ,  \chi_j\right) ~~~~
  \left[~~ {\bar\psi}^{1,\cdots,5}~~\right]  ~~~~~~~~~~~~~
\left({\bar\eta}_j\right)~~\nonumber\\
\left[\left(\matrix{4\cr
                                    0\cr}\right)+
\left(\matrix{4\cr
                                    2\cr}\right)+
\left(\matrix{4\cr
                                    4\cr}\right)\right]
& \left\{ 
\left(\matrix{2\cr
                                    0\cr}\right)\right.
 {\left[ \left(\matrix{5\cr
                                    0\cr}\right)+
\left(\matrix{5\cr
                                    2\cr}\right)+   
\left(\matrix{5\cr
                                    4\cr}\right)\right]
\left(\matrix{1\cr
                                    0\cr}\right)}\nonumber\\
&+~\left(\matrix{2\cr
                                   2\cr}\right)
{\left[\left(\matrix{5\cr
                                    1\cr}\right)+
\left(\matrix{5\cr
                                    3\cr}\right)+
\left(\matrix{5\cr
                                    5\cr}\right)\right]\left.
\left(\matrix{1\cr
                                    1\cr}\right)\right\}}
\label{spinor}
\end{eqnarray}
where
$4=\{y^3y^4,y^5y^6,{\bar y}^3{\bar y}^4,
{\bar y}^5{\bar y}^6\}$, $2=\{\psi^\mu,\chi^{12}\}$,
$5=\{{\bar\psi}^{1,\cdots,5}\}$ and $1=\{{\bar\eta}^1\}$.
The combinatorial factor counts the number of $\vert{-}\rangle$ in
a given state. The two terms in the curly brackets correspond to the two
components of a Weyl spinor.  The $10+1$ in the $27$ of $E_6$ are
obtained from the sector $b_j+\xi_1$.
The states which count the multiplicities of $E_6$ are the internal
fermionic states $\{y^{3,\cdots,6}\vert{\bar y}^{3,\cdots,6}\}$.
A similar result is
obtained for the sectors $b_2$ and $b_3$ with $\{y^{1,2},\omega^{5,6}
\vert{\bar y}^{1,2},{\bar\omega}^{5,6}\}$
and $\{\omega^{1,\cdots,4}\vert{\bar\omega}^{1,\cdots,4}\}$
respectively, which suggests that
these twelve states correspond to a six dimensional
compactified orbifold with Euler characteristic equal to 48.

The construction of the free fermionic models beyond the NAHE--set
entails the construction of additional boundary condition basis 
vectors and the associated one--loop GSO phases. Their
function is to reduce the number of generations and 
at the same time break the four dimensional gauge group.
In terms of the former the reduction is primarily by the
action on the set of internal world--sheet fermions
$\{y,\omega|{\bar y}, {\bar\omega}\}$.
As elaborated in the next section this 
set corresponds to the internal compactified manifold
and the action of the additional boundary condition basis
vectors on this set also breaks the gauge symmetries 
from the internal lattice enhancement. The later is 
obtained by the action on the gauge degrees of freedom
which correspond to the world--sheet fermions
$\{{\bar\psi}^{1,\cdots,5},{\bar\eta}^{1,\cdots,3},{\bar\phi}^{1,\cdots,8}\}$.
In the bosonic formulation this would correspond to
Wilson--line breaking of the gauge symmetries, hence for the purpose
of the reduction of the number of generations we
can focus on the assignment to the internal world--sheet fermions
$\{y,\omega|{\bar y},{\bar\omega}\}$.

We can therefore examine basis vectors that do not break the
gauge symmetries further, {\it i.e.} basis vectors of the 
form $b_j$, with 
$$\{\psi^\mu_{1,2}\chi_{j,j+1},(y,\omega|{\bar y},
{\bar\omega}),{\bar\psi}^{1,\cdots,5},{\bar\eta}_j\}=1$$
for some selection of $(y,\omega|{\bar y}, {\bar\omega})=1$
assignments such that the additional vectors
$b_j$ produce massless $SO(10)$ spinorials. We will refer
to such vectors as spinorial vectors. The additional
basis vectors $b_j$ can then produce chiral, or non--chiral, spectrum. 
The condition that the spectrum from a given such sector $b_j$
be chiral is that there exist another spinorial
vector, $b_i$, in the additive
group $\Xi$, such that the overlap between the
periodic fermions of the internal set $(y,\omega|{\bar y},
{\bar\omega})$ is empty, {\it i.e.}
\begin{equation}
\{b_j(y,\omega|{\bar y},{\bar\omega})\}\cap
\{b_i(y,\omega|{\bar y},{\bar\omega})\}\equiv\emptyset~.
\label{chiralitycondition}
\end{equation}
If there exists such a vector $b_i$ in the additive group then it
will induce a GSO projection that will select
the chiral states from the sector $b_j$. Interchangeably,
if such a vector does not exist, the states from the sector
$b_j$ will be non--chiral, ${\it i.e.}$ there will be an equal
number of $16$ and $\overline{16}$ or $27$ and $\overline{27}$.
For example, we note that for the NAHE--set basis vectors
the condition (\ref{chiralitycondition}) is satisfied.
Below I discuss the orbifold
correspondence of this condition.
The reduction to three generations in a specific model
is illustrated in table \ref{m278}.

\begin{eqnarray}
 &\begin{tabular}{c|c|ccc|c|ccc|c}
 ~ & $\psi^\mu$ & $\chi^{12}$ & $\chi^{34}$ & $\chi^{56}$ &
        $\bar{\psi}^{1,...,5} $ &
        $\bar{\eta}^1 $&
        $\bar{\eta}^2 $&
        $\bar{\eta}^3 $&
        $\bar{\phi}^{1,...,8} $\\
\hline
\hline
  ${\alpha}$  &  0 & 0&0&0 & 1~1~1~0~0 & 0 & 0 & 0 & 1~1~1~1~0~0~0~0 \\
  ${\beta}$   &  0 & 0&0&0 & 1~1~1~0~0 & 0 & 0 & 0 & 1~1~1~1~0~0~0~0 \\
  ${\gamma}$  &  0 & 0&0&0 &
		${1\over2}$~${1\over2}$~${1\over2}$~${1\over2}$~${1\over2}$
	      & ${1\over2}$ & ${1\over2}$ & ${1\over2}$ &
                ${1\over2}$~0~1~1~${1\over2}$~${1\over2}$~${1\over2}$~0 \\
\end{tabular}
   \nonumber\\
   ~  &  ~ \nonumber\\
   ~  &  ~ \nonumber\\
     &\begin{tabular}{c|c|c|c}
 ~&   $y^3{y}^6$
      $y^4{\bar y}^4$
      $y^5{\bar y}^5$
      ${\bar y}^3{\bar y}^6$
  &   $y^1{\omega}^5$
      $y^2{\bar y}^2$
      $\omega^6{\bar\omega}^6$
      ${\bar y}^1{\bar\omega}^5$
  &   $\omega^2{\omega}^4$
      $\omega^1{\bar\omega}^1$
      $\omega^3{\bar\omega}^3$
      ${\bar\omega}^2{\bar\omega}^4$ \\
\hline
\hline
$\alpha$ & 1 ~~~ 0 ~~~ 0 ~~~ 0  & 0 ~~~ 0 ~~~ 1 ~~~ 1  & 0 ~~~ 0 ~~~ 1 ~~~ 1
\\
$\beta$  & 0 ~~~ 0 ~~~ 1 ~~~ 1  & 1 ~~~ 0 ~~~ 0 ~~~ 0  & 0 ~~~ 1 ~~~ 0 ~~~ 1
\\
$\gamma$ & 0 ~~~ 1 ~~~ 0 ~~~ 1  & 0 ~~~ 1 ~~~ 0 ~~~ 1  & 1 ~~~ 0 ~~~ 0 ~~~ 0
\\
\end{tabular}
\label{m278}
\end{eqnarray}
In the realistic free fermionic models the vector $X$
is replaced by the vector $2\gamma$ in which $\{{\bar\psi}^{1,\cdots,5},
{\bar\eta}^1,{\bar\eta}^2,{\bar\eta}^3,{\bar\phi}^{1,\cdots,4}\}$
are periodic. This reflects the fact that these models
have (2,0) rather than (2,2) world-sheet supersymmetry.
At the level of the NAHE set we have 48 generations.
One half of the generations is projected because of the vector $2\gamma$.
Each of the three vectors in table \ref{m278}
acts nontrivially on the degenerate
vacuum of the fermionic states
$\{y,\omega\vert{\bar y},{\bar\omega}\}$ that are periodic in the
sectors $b_1$, $b_2$ and $b_3$ and reduces the combinatorial
factor of Eq. (\ref{spinor}) by a half.
Thus, we obtain one generation from each sector $b_1$, $b_2$ and $b_3$.

The geometrical interpretation of the basis vectors
beyond the NAHE set is facilitated by taking combinations of the
basis vectors in \ref{m278}, which entails choosing another set
to generate the same vacuum. The combinations
$\alpha+\beta$, $\alpha+\gamma$, $\alpha+\beta+\gamma$ produce
the following boundary conditions under the set of internal
real fermions

\begin{eqnarray}
     &\begin{tabular}{c|c|c|c}
 ~&   $y^3{y}^6$
      $y^4{\bar y}^4$
      $y^5{\bar y}^5$
      ${\bar y}^3{\bar y}^6$
  &   $y^1{\omega}^5$
      $y^2{\bar y}^2$
      $\omega^6{\bar\omega}^6$
      ${\bar y}^1{\bar\omega}^5$
  &   $\omega^2{\omega}^4$
      $\omega^1{\bar\omega}^1$
      $\omega^3{\bar\omega}^3$
      ${\bar\omega}^2{\bar\omega}^4$ \\
\hline
\hline
$\alpha+\beta$ 
& 1 ~~~ 0 ~~~ 1 ~~~ 1  & 1 ~~~ 0 ~~~ 1 ~~~ 1  & 0 ~~~ 1 ~~~ 1 ~~~ 0
\\
$\beta+\gamma$
& 0 ~~~ 1 ~~~ 1 ~~~ 0  & 1 ~~~ 1 ~~~ 0 ~~~ 1  & 1 ~~~ 1 ~~~ 0 ~~~ 1
\\
$\alpha+\beta+\gamma$
& 1 ~~~ 1 ~~~ 1 ~~~ 0  & 1 ~~~ 1 ~~~ 1 ~~~ 0  & 1 ~~~ 1 ~~~ 1 ~~~ 0
\\
\\
\end{tabular}
\label{m2782}
\end{eqnarray}

It is noted that the two combinations $\alpha+\beta$ and $\beta+\gamma$
are fully symmetric between the left and right movers, whereas the
third, $\alpha+\beta+\gamma$, is fully asymmetric.
The action of the first two combinations
on the compactified bosonic coordinates translates therefore to symmetric 
shifts. Thus, we see that reduction of the number of generations
is obtained by further action of fully symmetric shifts. 

Due to the presence of the third combination the situation, however, 
is more complicated. The third combination in \ref{m2782} is fully 
asymmetric between the left and right movers and therefore
does not have an obvious geometrical interpretation. In fact, 
a complete classification of all the possible 
$Z_2\times Z_2$ orbifold models with symmetric shifts on complex
tori, reveals that three generations are not obtained \cite{fknr}. 
Three generations are obtained in the free fermionic models
by the inclusion of the asymmetric shift. This observation
has profound implications on the type of geometries that
are related to the realistic string vacua, as well as on the
issue of moduli stabilization.

The same conclusion can also be obtained by using purely bosonic 
language. Starting with three complex tori parameterized by three
complex coordinates, the torus identification is given by (\ref{t2cube}). 
The symmetric shift action is 
$$z_i=z_i+{1\over 2}~~~{\rm and}~~~z_i=z_i+{\tau\over2}$$
and a given action may act on any number of the three tori.
The additional shifts may have the following actions:
\begin{eqnarray}
    {\rm freely ~acting}  & ~~\longrightarrow \oplus (h_{11}=h_{21}=0) \\
{\rm chiral ~preserving}  & \longrightarrow \oplus (h_{11}=h_{21}) \\
{\rm non ~freely ~acting} & \longrightarrow \oplus (h_{11}\ne h_{21})
\end{eqnarray}

In the first case one of the tori is always shifted and hence there
are no fixed points and the action is free. In the second case
we have tori above fixed points and all the other geometrical
identifications preserve the fixed tori. Since the contribution
of $T_2$ gives $\oplus h_{11}=\oplus h_{21}=1$
we have that this case preserves the chirality. In the third case
we have a situation that for a fixed torus we impose the identification
$z_k \leftrightarrow -z_k$. In this case the torus above the fixed
point degenerates to $P_1$, for which $\oplus h_{11}=1$,
$\oplus h_{21}=0$
and therefore this case adds to the net chirality. 
In ref. \cite{ron} we have classified all
the possible shifts on the three complex tori, and obtained
the same result. Three generations are not possible for $Z_2\times Z_2$
orbifolds of three complex tori, with purely symmetric shifts.

\section{Neutrino mass textures}\label{neutrino}

The neutrino sector of the Standard Model provides another piece
to the flavor enigma. Evidence for neutrino oscillations
steadily accumulated over the past few years, resulting in
compelling evidence for neutrino masses. This in turn points
to the augmentation of the Standard Model by the right--handed
neutrinos, and provides further evidence for the elegant
embedding of the Standard Model matter states, generation by generation,
in the 16 spinorial representation of $SO(10)$. However, in this
respect the new neutrino data raises further puzzles. The 
observation of a zenith angle dependence of $\nu_\mu$ 
from cosmic ray showers at super-Kamiokande \cite{superk}
provides strong evidence for oscillations in atmospheric neutrinos
with maximal $\nu_\mu\rightarrow\nu_\tau$ oscillations,
whereas the observations at the solar Sudbury Neutrino Observatory
(SNO) \cite{sno} and at the reactor KamLAND experiment \cite{kamland}
favor the large mixing angle MSW solution of the solar neutrino problem
\cite{MSW}. The recent data from the Wilkinson Microwave Anisotropy
Probe (WMAP) on cosmic microwave background anisotropies \cite{wmap},
combined with the 2 degree Field Galaxy Redshift Survey, CBI and ACBAR
\cite{galaxy},  restricts the amount of critical density 
attributed to relativistic neutrinos, and imposes that the
sum of the masses is smaller than $0.75$eV. 

While the Standard Model data strongly supports the incorporation of the 
Standard Model gauge and matter spectrum in representations
of larger gauge groups,
the flavor sector of the Standard Model provides further 
challenges. In the heavy generation the consistency
of the bottom--quark--tau lepton mass ratio with the 
experimental data arises due to the 
running of the strong gauge coupling.
The remaining flavor data, however, must
have its origin in a theory that incorporates gravity
into the picture. Most developed in this context are
the string theories that provide a viable perturbative framework
for quantum gravity. However, a new twist
of the puzzle arises due to the fact that while in
the quark sector we observe an hierarchical mass pattern
with suppressed mixing angles, the observations in the
neutrino sector are compatible with large mixing angles
that implies approximate mass degeneracy.

An elegant mechanism in the context of $SO(10)$ unification
to explain the large mixing in the neutrino sector was proposed
in ref. \cite{bajc}. However, this mechanism utilizes the 126 of
$SO(10)$, that does not arise in perturbative string theories \cite{dmr}.
On the other hand, string constructions
offer a solution to the proton longevity
problem. A doublet--triplet splitting mechanism is induced
when the $SO(10)$ symmetry is broken to $SO(6)\times SO(4)$
by Wilson--lines \cite{ps}. In the stringy doublet--triplet splitting
mechanism the color triplets are projected from the massless spectrum
and the doublets remain light. Additional symmetries that arise in the
string models may also explain the suppression of proton decay from
dimension four and gravity mediated operators. String constructions
also explain the existence of three generations in terms of the 
geometry of the compactified manifold. It is therefore 
important to seek other explanations for the origin of the 
discrepancy in the quark and lepton mass sectors.
An alternative possibility to the utilization of the 
126 in the seesaw mechanism is to use the nonrenormalizable term
$1616{\overline{16}}{\overline{16}}$. In this case the $B-L$ symmetry
is broken along a supersymmetric flat direction by the VEVs
of the neutral components of
$\langle16_H\rangle=\langle{\overline{16}}_H\rangle$,
where $16_H$ and ${\overline{16}}_H$ are two Higgs
multiplets, distinct from the three Standard Model generations. 
This term then induces the heavy Majorana mass term for the
right--handed neutrino. The contemporary studies of neutrino
masses in this context are based on this term. We will refer to this
as the ``one--step seesaw mechanism'' \cite{oneseesaw}. Similarly,
explorations in the context of type I string inspired models
also use the ``one--step seesaw mechanism'' \cite{king}.
In this talk we propose that the neutrino data points to the
role of $SO(10)$ singlet fields in the see--saw mass matrix.

The two--step seesaw mechanism utilizes the $SO(10)$ singlets
fields that are abundant in string models. An inspired version 
of this mechanism then takes the form 

\begin{equation}
  {\left(\matrix{
                   0  &     M_D &   0       \cr
                 M_D  &     0   & M_{\chi}  \cr
                   0  &  M_\chi & M_\phi    \cr
                }
   \right)}
\label{nmm}
\end{equation}

The left--handed Majorana mass matrix is given by

\begin{equation}
M_\nu=M_D M_\chi^{-1} M_\phi M_\chi^{-1} M_D^T~.
\label{mnu}
\end{equation}
 
The flavor structure of $m_{\nu_L}$ arises from $m_\phi$ and
can therefore account for the left--handed neutrino large 
while not disturbing the small mixing in the quark sector. 
This mechanism however requires the $\phi_i$ $SO(10)$ singlet
fields to exist at intermediate energies that presents a $\mu_\phi$
problem \cite{cfnu}.

\section{Conclusions and outlook}\label{uhecr}
I discussed in this paper the enterprise of string phenomenology
in light of LEP, KAMLAND and WMAP. These three experiments 
represent the reward of decades of dedicated
experimental efforts by numerous people, who contribute
anonymously to the accumulated scientific knowledge. Their
efforts should be saluted and are humbly and gratefully acknowledged.
Their results enable the theoretical pursuit of the proposition of
unification, string phenomenology, M-theory, and the like.
The particle experiments suggest two key ingredients that
a realistic vacuum should possess. The existence of
three generations together with their grand unification
embedding, most appealing in 16 spinorial representations of $SO(10)$.
The free fermionic models admit these two pivotal components.
It is of course not suggested that any of the contemporary
three generation models is the right one. That indeed would be folly.
Furthermore, it may also well be that at the end of the day
the free fermionic construction will not suffice to describe
the full details of the true vacuum. Nevertheless, it may still be the case
that the fermionic models are able to extract some key properties
of the true vacuum. It is then our task to try to isolate what those
key properties may be. It is proposed that the realistic free fermionic 
models suggest two such pivotal properties. The first is the relation
of the free fermionic point to the self--dual point under
T--duality. The second is the fact that the $Z_2$ orbifold twistings
that are utilized in the free fermionic models act on the coordinates
as real coordinates. This is furthermore compatible with the
necessity of introducing an asymmetric shift in the reduction to three
generations. Such a reduction may imply that the complex structure
of the Calabi--Yau threefold is necessarily broken. Hence providing
phenomenological guideline to the class of geometries that should be
sought to construct realistic string vacua. Furthermore, 
the asymmetric orbifold projection implies that some of the
moduli must be fixed. Incorporation of these observations in
the utilization of string dualities to study these vacua,
may provide further insight into their properties.
WMAP data paves the way for a new decade of exciting experimental
data that may yet challenge our perception of physical reality.
In the least it provides a more accurate account of the
energy composition of the universe, that the true string vacuum
should accommodate. The forthcoming years will provide further 
data with the UHECR experiments exploring the viability
of top--down models. The first priority of these experiments
is clearly to establish or refute the breaching of the GZK
cutoff, but the possibility that they may provide signals 
for new physics were entertained in ref. \cite{cafarella}.
Finally, the LHC will explore the electroweak symmetry breaking sector,
and subsequently
the TLC (The Linear Collider) will be vitally needed to provide
precision measurements.

%INDEX%%%%%%%%%%%%%%%%%%%%%%%%%%%%%%%%%%%%%%%%%%%%%%%%%%%%%%%%%%%%%%%
% Please check with the editor of your book whether he plans to
% include a "mutual" subject index - if so, please code your entries
% in the standard syntax. For your own purposes you may print your
% "personal" index by using the following commands:
%
%\clearpage
%\addcontentsline{toc}{section}{Index}
%\flushbottom
%\printindex
%%%%%%%%%%%%%%%%%%%%%%%%%%%%%%%%%%%%%%%%%%%%%%%%%%%%%%%%%%%%%%%%%%%%%

\end{document}